\documentclass[preprint,showpacs,preprintnumbers,amsmath,amssymb]{revtex4}

\usepackage{graphicx}
\usepackage{dcolumn}
\usepackage{bm}
\usepackage{epstopdf}

\newcount\driver
\newcount\bozza

scaled\magstep1                  
scaled\magstep1 \font\msytw=msbm10 scaled\magstep1
 
scaled\magstep1                 \font\indbf=cmbx10 scaled\magstep2

scaled \magstep2

{\count255=\time\divide\count255 by 60
\xdef\hourmin{\number\count255}
        \multiply\count255 by-60\advance\count255 by\time
   \xdef\hourmin{\hourmin:\ifnum\count255<10 0\fi\the\count255}}

\let\a=\alpha          \let\d=\delta     \let\e=\varepsilon
  \let\h=\eta     \let\th=\vartheta \let\k=\kappa     \let\l=\lambda
\let\m=\mu                          \let\r=\rho
             
\let\ps=\psi   \let\o=\omega     
\let\G=\Gamma \let\D=\Delta       \let\L=\Lambda

\def\EE{{\cal E}}\def\VV{{\cal V}}
\def\HH{{\cal H}}\def\WW{{\cal W}}

\def\RR{{\cal R}}\def\LL{{\cal L}}
\def\DD{{\cal D}}\def\AA{{\cal A}}

\def\qq{{\bf q}}\def\xx{{\bf x}}
   
\def\yy{{\bf y}}\def\kk{{\bf k}}\def\nn{{\bf n}}
\def\zz{{\bf z}}\def\uu{{\bf u}}

       \def\oo{{\underline \omega}}
\def\ee{{\underline \varepsilon}}  
          
          \def\ux{{\underline\xx}}

           \def\uy{{\underline\yy}}
\def\uz{{\underline \zz}}

\def\uu{\bf u}

\def\RRR{\hbox{\msytw R}}

        \def\EE{\hbox{\msytw E}}

\let\==\equiv

\let\io=\infty
\let\0=\noindent

\def\*{{\hfill\break\null\hfill\break}}

\def\tilde#1{{\widetilde #1}}

\def\lft{\left}
\def\rgt{\right}

\def\tende#1{\,\vtop{\ialign{##\crcr\rightarrowfill\crcr
             \noalign{\kern-1pt\nointerlineskip}
             \hskip3.pt${\scriptstyle #1}$\hskip3.pt\crcr}}\,}
\def\otto{\,{\kern-1.truept\leftarrow\kern-5.truept\to\kern-1.truept}\,}

\def\wh#1{\widehat{#1}}
\def\hat#1{\wh{#1}}
\def\sqt[#1]#2{\root #1\of {#2}}

\def\bp{{\bar \ps}}

\def\EE{{\cal E}}\def\VV{{\cal V}}
\def\HH{{\cal H}}\def\WW{{\cal W}}

\def\RR{{\cal R}}\def\LL{{\cal L}}
\def\DD{{\cal D}}\def\AA{{\cal A}}

\def\T#1{{#1_{\kern-3pt\lower7pt\hbox{$\widetilde{}$}}\kern3pt}}
\def\VVV#1{{\underline #1}_{\kern-3pt
\lower7pt\hbox{$\widetilde{}$}}\kern3pt\,}
\def\W#1{#1_{\kern-3pt\lower7.5pt\hbox{$\widetilde{}$}}\kern2pt\,}

\def\indica{\leaders \hbox to 0.5cm{\hss.\hss}\hfill}
\def\guida{\leaders\hbox to 1em{\hss.\hss}\hfill}
\mathchardef\oo= "0521

\def\qq{{\bf q}}\def\xx{{\bf x}}
\def\yy{{\bf y}}\def\kk{{\bf k}}\def\nn{{\bf n}}
\def\zz{{\bf z}}\def\uu{{\bf u}}

\def\oo{{\underline \omega}}

\def\qed{\raise1pt\hbox{\vrule height5pt width5pt depth0pt}}
 
  \def\bp{{\bar p}} 

\def\indic{\hbox{\raise-2pt \hbox{\indbf 1}}}

\def\RRR{\hbox{\msytw R}}

\def\ins#1#2#3{\vbox to0pt{\kern-#2 \hbox{\kern#1 #3}\vss}\nointerlineskip}

\newdimen\xshift \newdimen\xwidth \newdimen\yshift
\newcount\griglia

\def\insertplot#1#2#3#4#5#6{%
\xwidth=#1pt \xshift=\hsize \advance\xshift by-\xwidth \divide\xshift by 2%
\begin{figure}[ht]
\vspace{#2pt} \hspace{\xshift}
\begin{minipage}{#1pt}
#3 \ifnum\driver=1 \griglia=#6
\ifnum\griglia=1 \openout13=griglia.ps \write13{gsave .2
setlinewidth} \write13{0 10 #1 {dup 0 moveto #2 lineto } for}
\write13{0 10 #2 {dup 0 exch moveto #1 exch lineto } for}
\write13{stroke} \write13{.5 setlinewidth} \write13{0 50 #1 {dup 0
moveto #2 lineto } for} \write13{0 50 #2 {dup 0 exch moveto #1
exch lineto } for} \write13{stroke grestore} \closeout13
\includegraphics{griglia.ps} \fi
\includegraphics{#4.ps}\fi%
\ifnum\driver=2 \fi
\end{minipage}
\caption{#5}
\end{figure}
}

\newdimen\shift \shift=-1.5truecm
\def\lb#1{%
\ifnum\bozza=1
\label{#1}\rlap{\hbox{\hskip\shift$\scriptstyle#1$}}
\else\label{#1} \fi}

\def\be{\begin{equation}}
\def\ee{\end{equation}}
\def\bea{\begin{eqnarray}}\def\eea{\end{eqnarray}}
\def\bean{\begin{eqnarray*}}\def\eean{\end{eqnarray*}}
\def\bfr{\begin{flushright}}\def\efr{\end{flushright}}
\def\bc{\begin{center}}\def\ec{\end{center}}
\def\bal{\begin{align}}\def\eal{\end{align}}
\def\ba#1{\begin{array}{#1}} \def\ea{\end{array}}
\def\bd{\begin{description}}\def\ed{\end{description}}
\def\bv{\begin{verbatim}}\def\ev{\end{verbatim}}
\def\nn{\nonumber}
\def\Halmos{\hfill\vrule height10pt width4pt depth2pt \par\hbox to \hsize{}}
\def\pref#1{(\ref{#1})}

\def\ins#1#2#3{\vbox to0pt{\kern-#2 \hbox{\kern#1 #3}\vss}\nointerlineskip}

\newdimen\xshift \newdimen\xwidth \newdimen\yshift
\newcount\griglia

\def\insertplot#1#2#3#4#5#6{%
\xwidth=#1pt \xshift=\hsize \advance\xshift by-\xwidth \divide\xshift by 2%
\begin{figure}[ht]
\vspace{#2pt} \hspace{\xshift}
\begin{minipage}{#1pt}
#3 \ifnum\driver=1 \griglia=#6
\ifnum\griglia=1 \openout13=griglia.ps \write13{gsave .2
setlinewidth} \write13{0 10 #1 {dup 0 moveto #2 lineto } for}
\write13{0 10 #2 {dup 0 exch moveto #1 exch lineto } for}
\write13{stroke} \write13{.5 setlinewidth} \write13{0 50 #1 {dup 0
moveto #2 lineto } for} \write13{0 50 #2 {dup 0 exch moveto #1
exch lineto } for} \write13{stroke grestore} \closeout13
\includegraphics{griglia.ps} \fi
\includegraphics{#4.ps}\fi%
\ifnum\driver=2 \fi
\end{minipage}
\caption{#5}
\end{figure}
}

\newdimen\shift \shift=-1.5truecm
\def\lb#1{%
\label{#1}\rlap{\hbox{\hskip\shift$\scriptstyle#1$}}
\else\label{#1} \fi}

\def\be{\begin{equation}}
\def\ee{\end{equation}}
\def\bea{\begin{eqnarray}}\def\eea{\end{eqnarray}}
\def\bean{\begin{eqnarray*}}\def\eean{\end{eqnarray*}}
\def\bfr{\begin{flushright}}\def\efr{\end{flushright}}
\def\bc{\begin{center}}\def\ec{\end{center}}
\def\bal{\begin{align}}\def\eal{\end{align}}
\def\ba#1{\begin{array}{#1}} \def\ea{\end{array}}
\def\bd{\begin{description}}\def\ed{\end{description}}
\def\bv{\begin{verbatim}}\def\ev{\end{verbatim}}
\def\nn{\nonumber}
\def\Halmos{\hfill\vrule height10pt width4pt depth2pt \par\hbox to \hsize{}}
\def\pref#1{(\ref{#1})}

\usepackage{amsmath}
\usepackage{amsfonts}
\usepackage{amssymb}
\usepackage{slashbox}
\usepackage{epstopdf}

\font\msytw=msbm9 scaled\magstep1 
scaled\magstep1

\let\a=\alpha     \let\d=\delta
\let\e=\varepsilon
  \let\h=\eta   \let\th=\theta \let\k=\kappa \let\l=\lambda
\let\m=\mu                 \let\r=\rho
     
\let\ps=\Psi   \let\o=\omega
\let\G=\Gamma \let\D=\Delta  \let\L=\Lambda

\def\EE{{\cal E}} \def\VV{{\cal V}}
 \def\WW{{\cal W}}
 
\def\RR{{\cal R}}\def\LL{{\cal L}}  
\def\DD{{\cal D}}

\def\qq{{\bf q}} 
 \def\xx{{\bf x}} \def\yy{{\bf y}} \def\zz{{\bf z}}
\def\kk{{\bf k}}

\def\nn{\nonumber}

\def\RRR{\hbox{\msytw R}}

\def\\{\hfill\break}
\def\={:=}
\let\io=\infty
\let\0=\noindent

\def\tende#1{\,\vtop{\ialign{##\crcr\rightarrowfill\crcr\noalign{\kern-1pt
    \nointerlineskip} \hskip3.pt${\scriptstyle #1}$\hskip3.pt\crcr}}\,}
\def\otto{\,{\kern-1.truept\leftarrow\kern-5.truept\to\kern-1.truept}\,}

\def\wh{\widehat}
\def\to{\rightarrow}

\def\qed{\hfill\raise1pt\hbox{\vrule height5pt width5pt depth0pt}}

\def\be{\begin{equation}}
\def\ee{\end{equation}}
\def\bp{\begin{pmatrix}}
\def\ep{\end{pmatrix}}
\def\bea{\begin{eqnarray}}
\def\eea{\end{eqnarray}}
\def\nn{\nonumber}
\def\pref#1{(\ref{#1})}

\def\lb{\label}

\begin{document}

\title{Ward Identities and chiral anomalies for coupled fermionic chains}

\author{L.C. Costa}
\affiliation{%
Centro de Ci\^encias Naturais e Humanas,
Universidade Federal do ABC, 09210-170, Santo Andr\'e, Brazil
}
\author{
A.Ferraz}
\affiliation{%
International Institute of Physics and Dept. of Theoretical and Experimental physics, Universidade Federal do Rio Grande do Norte,
59012-970, Natal, Brazil.}

\author{Vieri Mastropietro}

\affiliation{Dipartimento di Matematica F. Enriques,
Universit\'a di Milano, Via C. Saldini 50, 20133, Milano, Italy
}

\begin{abstract} Coupled fermionic chains
are usually described by an effective model
written in terms of bonding and anti-bonding spinless fields with linear dispersion
in the vicinities of the respective Fermi points. We derive for the first time
exact Ward Identities (WI) for this model, proving the
existence of chiral anomalies which
verify the Adler-Bardeen
non-renormalization property. Such WI are expected to play a crucial role in the understanding of the thermodynamic properties
of the system.
Our results are non-perturbative and are
obtained analyzing Grassmann functional integrals by means of Constructive Quantum Field Theory methods.
\end{abstract}

\maketitle

\section{Introduction}

Coupled fermionic chains have been extensively analyzed in the last years, ever since the suggestion
\cite{AN87}-\cite{CL95} that they are related to the
physics of high $T_c$ superconductors. In spite of that, their properties in the thermodynamic limit are still largely unknown.
The Hamiltonian of two spinless interacting fermionic chains coupled by a
hopping term is
\be H=H^{(1)}+H^{(2)}+t_\perp \sum_x [a^+_{x,1} a^{-}_{x,2}+a^+_{x,2}
a^{-}_{x,1}]\label{cc} \ee
where $a^\pm_{x,1}, a^\pm_{x,2}$ are fermionic creation or
annihilation operators, $x=1,2,..,L$ and \be H^{(i)}=
\sum_{x=1}^{L-1} [-{1\over 2} ( a^+_{x+1,i}a^-_{x,i}+a^+_{x,i}a^-_{x+1,i}) - \m
a^+_{x,i}a^-_{x,i}]-\l\sum_{x,y=1}^{L-1}
v(x-y)a^+_{x,i}a^-_{x,i} a^+_{y,i}a^-_{y,i} \ee
with $i= 1, 2$ and where $|v(x)|\le e^{-\k |x|}$, $v(x)=v(-x)$.
%
%
When $t_\perp=0$
the system reduces to two uncoupled chains and for such a case the
{\it Luttinger liquid behavior} (in the sense of \cite{Ha1}) has been rigorously
established in \cite{BM0},\cite{BM1},\cite{BM2}, using Renormalization Group methods combined with Ward the Identities.
The interchain hopping is however a relevant interaction which can produce a radically different behavior
in such a system.

The analysis
of the single chain in  \cite{BM0}, \cite{BM1},\cite{BM2} succeeds in implementing in
a rigorous way the deep idea, due to Tomonaga \cite{To}, of the {\it emergence} in 1d metals
of a low energy description in terms of an effective Quantum Field Theory, which in the case of spinless 1d metal turns out to be the Thirring or the Luttinger model \cite{L}. This was used to establish
asymptotic Ward Identities which are essential for the non-perturbative Renormalization Group analysis.
It is therefore natural to consider
an effective Quantum  Field Theory description for the coupled chain model \pref{cc} as the starting point of the analysis,
and this was done in  \cite{2a,2}.
The effective low energy description for the coupled chain model
\pref{cc} is obtained introducing
the {\it bonding} and {\it antibonding} fermionic field  operators
\be
b_{x,a}^\pm={1\over\sqrt{2}}(a^\pm_{x,1}-a^\pm_{x,2}) , \quad \quad
b_{x,b}^\pm={1\over\sqrt{2}}(a^\pm_{x,1}+a^\pm_{x,2})\label{asas} \ee
In terms of these variables
the Hamiltonian \pref{cc} can be written as
\bea
&&H=H_0+V\label{cc1}\\
&&H_0={1\over 2}\sum_{j=a,b}\sum_x [-b^+_{x+1,j}b^-_{x,j}-b^+_{x,j}b^-_{x+1,j}-\m_j b^+_{x,j}b^-_{x,j}]
\nn\\
&&V=-{\l\over 2}\sum_{j=a,b}\sum_{x,y} v(x-y)[b^+_{x,j} b^-_{x,j} b^+_{y,j} b^-_{y,j}+
b^+_{x,j} b^-_{x,j} b^+_{y,-j} b^-_{y,-j}+
b^+_{x,j} b^-_{x,-j} b^+_{y,j} b^-_{y,-j}
+b^+_{x,j} b^-_{x,-j} b^+_{y,-j} b^-_{y,j}]\nn
\eea
with $\m_a=(\m+t_\perp)$, $\m_b=(\m-t_\perp)$ and we used the convention $-a=b, -b=a$.
Therefore the  model \pref{cc}, describing two interacting chains coupled by a quadratic hopping term,
appears to be equivalent, with this change of variables, to
a couple of fermionic chains with different chemical potentials and which are coupled by a quartic interaction.

The two possible descriptions of the two chains model are convenient in different
energy (or temperature) regimes. Indeed,
as shown in \cite{M0},   for $\l$ and $t_\perp$ small enough, and temperatures such that
\be
T\ge |t_\perp|^{1\over 1-\h}\label{cond}
\ee
with $\h=a\l^2+O(\l^3)$ and $a>0$ , the behavior of the two coupled chains correlations is essentially the same of the uncoupled ones up to small corrections. This fact
has important consequences for the transverse conductivity between chains. This result is proved using a (renormalized) expansion in $t_\perp$ (using the representation \pref{cc})
which is convergent for small $t_\perp$ and under the condition \pref{cond}; the temperature condition ensures that no other marginal quartic couplings appear in addition
to the ones associated with the single chain. The main effect of the interaction is to produce a non trivial renormalization
of the hopping, as it is clear from \pref{cond}. In particular the interaction strongly reduces the interchain hopping when $|\l|>>|t_\perp|$.

In the lower temperature regime $T< |t_\perp|^{1\over 1-\h}$ there is however no hope that an expansion in $t_\perp$ should work, and it is better to rely on the representation \pref{cc1}.
The behavior of the system
in this second region is much less understood and the representation \pref{cc1} is likely to be much more adequate.
As we said earlier, it is natural to follow what was done in the single chain and introduce an effective Quantum Field Theory description, following the same steps which lead
to the introduction of the Luttinger model. The model \pref{cc1} describes two kinds of fermions with Fermi surface characterized by two Fermi points $\pm p_F^{(j)}$
(the magnitudes of the Fermi points will be in general different with respect to its bare values at $\l=0$, and
it will depend from the {\it renormalized} hopping)
and close to such points the dispersion relation is approximately linear.
This suggests that, as in \cite{To}, each of the bonding or antibonding fermions
can be described
in terms of excitations $e^{\pm i\o  p_F^{(j)}x} b^\pm_{x,\o,j}$
, which take place close to each one of the Fermi points; an extra index $\o=\pm$, in addition to $j=a,b$,
denotes the right or left Fermi point and $x$ becomes a continuum variable. Another important assumption is that the
fermions with index $\o=\pm$ have a linear "relativistic" dispersion relation $\o k$ (the momenta are measured from the Fermi points, and this explains the oscillating factor $e^{\pm i\o  p_F^{(j)}x})$. This should require a momentum cut-off
(the dispersion relation is approximately linear only close to the Fermi points) but one can hope that the long distance properties
are the same even without it. This however is not a trivial issue since the linear dispersion relation can produce
ultraviolet divergences which are typical in Quantum Field Theory, which are of course absent in the lattice
model \pref{cc}.
%
%

The interaction of the effective model can be obtained from \pref{cc1}
replacing each field $b^\pm_{x,j}$ with  $\sum_{\o=\pm }e^{\pm i\o  p_F^{(j)}x} b^\pm_{x,\o,j}$,
considering $x$ a continuum variable. One obtains a number of terms but some of them appears to be negligible (again,
one follows here the derivation in \cite{To} of the Luttinger model). In particular several terms
(for instance
$b^+_{\o,j}b^-_{\o,j} b^+_{\o,j} b^-_{-\o,j}$), in which the sum of the momenta measured from the Fermi points is non vanishing and is of order $O(1)$,
can be promptly neglected. One keeps only the terms such that the sum of the momenta is either zero or  $O(p_F^{(a)}-p_F^{(b)})$.
Note that in the second case
such difference is small when $t_\perp$ is small and therefore one cannot neglect them safely.

The above considerations lead to an effective model
(essentially introduced in  \cite{2a,2}) which we will describe in a precise way next.
Such a model is just the analogue of the Luttinger model for the single chain model, and it is interesting
to study its properties as a preliminary step for the analysis of the more realistic model \pref{cc}.
However,  while the Luttinger model, resulting from the low energy description
of the single chain problem, is exactly solvable \cite{ML}, the effective model for the two chain problem appears {\it not} to be solvable
and consequently it is much more difficult to analyze.
This is the reason why the thermodynamic properties
are still controversial, despite an intense investigation along the years, see {\it e.g.}
\cite{2a,2,0,0a,FA93,11a,12a,13a,7,15a,LE05,M0,17a,TH12}. In particular bosonization, which was used in \cite{ML}
for the solution of the Luttinger model, has not been useful in this case, so far.

The effective two chain model
is expressed in terms of
Grassmann variables.
Given $L>0$ we consider the set $\DD$ of space-time
momenta $\kk=(k,k_0)$, with $k={2\pi\over L}(n+{1\over 2})$ and
$k_0={2\pi\over L}(n_0+{1\over 2})$; for each $\kk\in \DD$ we
associate eight {\it Grassmann variables} (sometimes also called
fields) $\hat\psi^+_{\kk,\o,j},\hat\psi^-_{\kk,\o,j}$ with
$\o=(+,-)$ a {\it chiral} or Fermi point index and $j=a,b$ the, {\it band}
index. We also define
\be \psi^\pm_{\xx,\o,j}={1\over L^2}\sum_{\kk\in \DD} e^{\pm
i\kk\xx} \hat\psi^\pm_{\kk,\o,j}\ee
where $\xx=(x,x_0)\in \L$, $\L$ being a bidimensional square torus with side $L$.
Note that $x$ is now a continuum variable and not a lattice one as in \pref{cc},\pref{cc1}.
The {\it Generating Functional} of the effective model is the following
Grassmann integral
\bea && e^{W_{N,L}(\h,J)} = \int\!
P(d\psi)\; \exp \left\{ -V(\psi)
+ \sum_{\o,j} \int \!d\xx\; J_{\xx,\o}
\r_{\xx,\o,j}\right.\nn\\
&& + \left. \sum_{\o,j} \int\! d\xx\; \lft[\psi^{+}_{\xx,\o,j}
\h^-_{\xx,\o,j} + \h^+_{\xx,\o,j}
\psi^{-}_{\xx,\o,j}\rgt]\right\}\;\label{caro}\eea
where $\r_{\xx,\o,j}=\psi^{+}_{\xx,\o,j} \psi^{-}_{\xx,\o,j}$,
$\h,J$ are (respectively commuting and anticommuting) {\it
external fields}, $P(d\psi)$ is the {\it Gaussian
Grassmann measure} with propagator
$\d_{\o,\o'}\d_{j,j'}g_{\o}(\xx-\yy)$ with
\be g_{\o}(\xx-\yy)={1\over
L^2}\sum_{\kk\in \DD}e^{i\kk(\xx-\yy)}{\chi^\e_{N}(\kk)\over -ik_0+\o  k}\label{pro} \ee
where
$\chi^\e_{N}(\kk)$ is a cut-off function nonvanishing
for all $\kk$, depending on a small positive parameter $\e$, and
reducing itself, as $\e\to 0$, to a compact support function
$\chi_N(\kk)=\bar\chi(2^{-N} |\kk|)$
with  $\bar\chi(t)=1$ for $0\le t\le
1$ and vanishing for
$t\ge 2$. The interaction is given by
\begin{eqnarray} &&V(\psi) =\label{int1}
\sum\limits_{j,\o} \int d\xx d\yy v(\xx-\yy)
\bigg\{g_{0}\psi^+_{\xx,\o,j}\psi^-_{\xx,\o,j}
\psi^+_{\yy,-\o,j}\psi^-_{\yy,-\o,j}+\nn\\
&&g_{f} \psi^+_{\xx,\o,j}
\psi^-_{\xx,\o,j}\psi^+_{\yy,-\o,-j}
\psi^-_{\yy,-\o,-j}+g_{u}\psi^+_{\xx,\o,j}
\psi^-_{\xx,\o,-j}\psi^+_{\yy,-\o,j}
\psi^-_{\yy,-\o,-j}+\nn\\
&&g_{bs}\psi^+_{\xx,\o,j}
\psi^-_{\xx,\o,-j}\psi^+_{\yy,-\o,-j}\psi^-_{\yy,-\o,j}+g_{4} \psi^+_{\xx,\o,j}
\psi^-_{\xx,\o,j}\psi^+_{\yy,\o,-j}
\psi^-_{\yy,\o,-j}+\\
&&\tilde g_4
\psi^+_{\xx,\o,j}
\psi^-_{\xx,\o,j}\psi^+_{\yy,\o,j}
\psi^-_{\yy,\o,j}
\bigg\}, \nn
\end{eqnarray}
%
%
with the convention $-a=b, -b=a$,
 $|v(\xx-\xx')|\le e^{-\k |\xx-\xx'|}$, $\hat v(0)=1$; the first and second terms are called
{\it forward} interactions, the third {\it umklapp} and the fourth {\it backscattering}.
The {\it Schwinger functions} are determined by the functional derivatives
of the generating functional; for instance
\bea
&&<\psi^{-}_{\xx,\o,j}\psi^{+}_{\yy,\o,j}>_{N,L,\e}={\partial^{2}W_{N,L,\e}(\h,J)
\over\partial\h^{+}_{\xx,\o,j}\partial\h^{-}_{\yy,\o,j}}]\Big|_{0,0}\nn\\
&&<\r_{\zz,\o',j'};\psi^{-}_{\xx,\o,j}\psi^{+}_{\yy,\o,j}>_{N,L,\e} ={\partial^{3}W_{N,L}(\h,J)
\over\partial J_{\zz,\o',j'}\partial\h^{+}_{\xx,\o,j}\partial\h^{-}_{\yy,\o,j}}\Big |_{0,0}
\eea
and we denote $\lim_{\e\to 0}<.>_{L,N,\e}=<.>_{L,N}$.

In order to have a well defined functional integral we make use of an infrared cut-off (the finite volume $\L$)
and an ultraviolet cut-off $2^N$, which will be removed eventually when we take
the limit $N\to\io$. The only role of the parameter $\e$ in the
definition of the cut-off function is in making the derivation of Ward Identities easier. The limit $\e\to 0$ will be performed first. The non local nature of the interaction has also the role of an ultraviolet cut-off;
we have no need of
{\it counterterms} unbounded in $N$ to perform the ultraviolet limit.

Note finally that the interaction is invariant under the transformations
\be\psi^\pm_{\xx,\o,j}\to e^{\pm i\a_{\xx,\o}}\psi^\pm_{\xx,\o,j}\label{zzz}\ee
where the phase is  $\o$-dependent but not $j$-dependent. In the special case $g_u=g_{bs}=0$ the interaction is invariant under the transformation
\be\psi^\pm_{\xx,\o,j}\to e^{\pm i\a_{\xx,\o,j}}\psi^\pm_{\xx,\o,j}\label{zzz1}\ee
with the phase being both $\o$ and $j$-dependent. Our aim is to derive a set of Ward Identities associated with the invariance \pref{zzz}.

\section{Ward Identities and chiral anomalies}

By performing in the
functional integral \pref{caro} the change of variables \pref{zzz} one gets
\bea && e^{W_{N,L}(\h,J)} = \int\!
P(d\psi)\; e^{-V(\psi)-\sum_{\o,j} \int d\xx \psi^+_{\xx,\o,j}[e^{i\a_{\xx,\o}}D_\xx e^{-i\a_{\xx,\o}}-D_\xx]
\psi^-_{\xx,\o,j}}\\
&&e^{\sum_{\o,j} \int \!d\xx\; [ J_{\xx,\o}
\psi^{+}_{\xx,\o,j} \psi^{-}_{\xx,\o,j}+ e^{i\a_{\o,\xx}}\psi^{+}_{\xx,\o,j}
\h^-_{\xx,\o,j} + e^{-i\a_{\o,\xx}}\h^+_{\xx,\o,j}
\psi^{-}_{\xx,\o,j}] }\;\label{caro1}\nn\eea
where
\be
D_\xx \psi_{\xx,\o,j}={1\over L^2}\sum_\kk e^{i\kk\xx} (\chi^\e_N(\kk))^{-1}D_\o(\kk)\psi^+_{\kk,\o,j}\;.
\ee
and $D_{\o}(\kk)=-i k_0+\o k$.
Here we have taken into account that the Jacobian of the transformation is $1$ (this is true only if the change of variables is done at non vanishing $\e$), see \S 2 of \cite{BM1} for more details (see also
\cite{K}).
By performing derivatives with respect to  $\hat \a_{\qq,\o},\hat\h^+_{\kk,\o',j'},
\hat\h^-_{\kk+\qq,\o',j'}$, we get the following identity (written directly in the limit $\e\to 0$):
\bea &&\sum_j D_{\o}(\qq)<\hat\r_{\qq,\o,j};\hat\psi^-_{\kk,\o',j'}
\hat\psi^+_{\kk+\qq,\o',j'}>_{N,L}=\label{h1a}   \\
&&\d_{\o,\o'}[<\hat\psi^-_{\kk+\qq,\o',j'}
\hat\psi^+_{\kk+\qq,\o',j'}>_{N,L}- <\hat\psi^-_{\kk,\o',j'}
\hat\psi^+_{\kk,\o',j'}>_{N,L}]+\hat\D_{N,L}(\kk,\kk+\qq)\nn \eea
where
\be
\hat\D(\kk,\kk+\qq)={1\over L^2}\sum_{\kk'} C^N_{\o}(\kk')<\hat\psi^+_{\kk',\o,j}
\hat\psi^+_{\kk'+\qq,\o,j};\hat\psi^-_{\kk,\o',j'}
\hat\psi^+_{\kk+\qq,\o',j'}>
\ee
and
\be C^N_{\o}(\kk,\qq)=[(\chi^\e_{N}(
\kk+\qq) )^{-1}- 1]
D_{\o}(\kk+\qq) - [(\chi^\e_{N}(
\kk)^{-1} - 1]
D_{\o}(\kk)\label{fonddd1}. \ee
The last term in \pref{h1a}, namely $\hat\D_{N,L}(\kk,\kk+\qq)$, can be considered a
correction term due to the fact that the momentum cut-off breaks the gauge invariance of the theory. It is a rather complicate expression and it is not a Schwinger function.
One may suspect that,
when the ultraviolet cut-off is removed, that is the limit $N\to\io$ is taken, than such correction term is nullified. This is {\it not} what happens, as $\hat\D_{N,L}$ gives a non-vanishing contribution
in the limit $N\to\io$. This provides an example of a {\it quantum anomaly}, and
is the content of the following Theorem, which is the main result of this paper. For convenience
we will use the notation $\vec g=g_0,g_f,g_u,g_{bs},g_4,\tilde g_4$.
\vskip.3cm
{\bf Theorem.} {\it Given the generating functional \pref{caro}, there exists $\e_L$ such that, for $|\vec g|\le \e_L$
the limits \bea
&&\lim_{N\to\io}<\hat\r_{\qq,j,\o};\hat\psi^-_{\kk,\o',j'}
\hat\psi^+_{\kk+\qq,j',\o'}>_{N,L}=<\hat\r_{\qq,\o,j};\hat\psi^-_{\kk,\o',j'}
\hat\psi^+_{\kk+\qq,\o',j'}>_{L}\nn\\
&&\lim_{N\to\io}<\hat\psi^-_{\kk,\o',j'}
\hat\psi^+_{\kk,\o',j'}>_{N,L}=<\hat\psi^-_{\kk,\o',j'}
\hat\psi^+_{\kk,\o',j'}>_{N}\label{h3}
\eea
exist and verify the Ward Identity \pref{h1a} with
\bea
&&\lim_{N\to\io}\hat\D_{L,N}(\kk,\kk+\qq)=\label{ll22}\\
&&\sum_{j=a,b}\{{g_0\over 4\pi }D_{-\o}(\qq)<\hat\r_{\qq,-\o,j};\hat\psi^-_{\kk,\o',j'}
\hat\psi^+_{\kk+\qq,\o',j'}>_{L}+{g_f\over 4\pi}D_{-\o}(\qq)<\hat\r_{\qq,-j,-\o};\hat\psi^-_{\kk,\o',j'}
\hat\psi^+_{\kk+\qq,\o',j''}>_{L}\nn\\
&&+{g_4\over 4\pi }D_{-\o}(\qq)<\hat\r_{\qq,\o,-j};\hat\psi^-_{\kk,\o',j'}
\hat\psi^+_{\kk+\qq,\o',j'}>_{L}+{\tilde g_4\over 4\pi }D_{-\o}(\qq)<\hat\r_{\qq,\o,j};
\hat\psi^-_{\kk,\o',j'}
\hat\psi^+_{\kk+\qq,\o',j'}>_{L}\}\nn
\eea
with the notation $-a=b, -b=a$ as before.
}
\vskip.3cm

{\bf Remarks}
\begin{enumerate}
\item In the limit $N\to\io$ the correction term to the Ward Identities is non vanishing,
but its form radically simplifies and it can
be expressed as sum over Schwinger functions.
The coefficients multiplying the correlations in the r.h.s.
of \pref{ll22} are called {\it anomalies}.
Remarkably, the anomalies are {\it linear}
in the coupling. That is, all the possible higher order corrections give no contribution.
Such a property is the {\it non perturbative} analogue of the
{\it anomaly non renormalization} in QFT, established by Adler and Bardeen for
$(3+1)$ dimensional Quantum Electrodynamics in \cite{AB}
at all orders in perturbation theory
and extended to QFT models
in $(1+1)$ dimensions in \cite{GR}.
\item The only contribution to the anomalies comes from the
interactions verifying \pref{zzz1}, that is $g_0,g_f,g_4,\tilde g_4$;
remarkably in a model with only the couplings $g_{bs}, g_u$ the WI has no anomalies.
\item If we set $g_u=g_{bs}=0$ and we consider the Hamiltonian analogue of the
effective two chain model, as defined in  \cite{2a,2}
(in which
a cut-off only on the spatial part of the momenta is implicit),
one gets a system which can be solved by bosonization, and one can derive
for such system a WI with an anomalous term coinciding with ours.
This is however true {\it only} if
$g_{u}=g_{bs}=0$, and
there is so far no way of
deriving the WI by bosonization for the full interaction
\pref{int1}.
\item  The Theorem is proved using the methods
introduced in \cite{M1} for massless $QED_{1+1}$. There is however a crucial difference with respect of such case; massless $QED_{1+1}$
is closely related to a solvable model (the Thirring model) and as a consequence
of that one can take safely the infinite volume limit in this case.  The main novelty of the present case is that
the system has not an underlying solvable model and
the flow of the running coupling constants is unbounded in the infrared.
Despite this fact, we can prove
that an  {\it exact} WI is true when the ultraviolet cut-off is removed, even at finite volume, with the resulting chiral anomaly verifying the non-renormalization property.
\item In the Theorem we have written only the WI corresponding to the vertex, but
one could derive an infinite number of WI by an easy extension of the analysis explained below.
\item Such WI are among the very few properties one can derive
rigorously for the effective two chain model, and we believe that they will play an important role in the understanding of its (still largely unknown)
properties in the thermodynamic limit, in analogy to what happen for the single chain problem.
\item A similar WI can be derived also for the density correlations and, as briefly discussed in Appendix A,
this
allows us to deduce the exact form of the total density correlation (see (66) below), which is identical to the
non interacting one
up to a renormalization of the amplitude and of the velocity. Physically, this simply means that the
total density excitations remain gapless , as in a Luttinger liquid (of course a gap could appear in other density excitations) .
\item Even when a full understanding of the effective model
will be reached, it remains an open problem the relation with the original model \pref{cc}. How
the parameters of the effective model relate to the parameters in \pref{cc}
is also an open question. Hopefully the rigorous Renormalization Group techniques, which could prove the emergence of the Luttinger model in the single chain model, will provide
the answer to this question.

\end{enumerate}

The rest of the paper is organized in the following way. In \S 3 we will prove the main
Theorem above. In App. A we discuss the dependence on our results on
the cut-off function and the relation with bosonization methods.
Finally, in App. B we
explicitly check
the WI \pref{h1a},\pref{ll22} up to second order in perturbation theory.

\section{Proof}

\subsection{Multiscale decomposition of the generating functional}

The proof is based on \cite{M1}, where an effective model for a single chain was analyzed, in which the fermions
have no $j$ index and the interaction contains only one term. We will focus here
on the extensions and modifications necessary to the present case, referring to
that paper for several technical results.

We write
the cut-off function as (directly in the limit $\e\to 0$)
\be \chi_N(\kk)=\sum_{k=-\io}^{N} f_k(\kk)\label{cf}, \ee
where $f_k(\kk)=\bar\chi(2^{-k} |\kk|)-\bar\chi(2^{-k+1} |\kk|)$
is a smooth function of $|\kk|$
which is non vanishing only for  $2^{k-1}\le |\kk|\le 2^{k+1}$.
%
Note that since $|k_0|$ and $|k|$ are greater than ${\pi\over L}$
there exists a scale $h_L$ such that $f_k(\kk)=0$ for $k\le
h_L$ with $-h_L=O(\log L)$; we can then write
\be g_{\o}(\xx-\yy)=\sum_{k=h_L}^{N} ({1\over L})^2\sum_\kk e^{i\kk(\xx-\yy)}
{f_k( \kk)\over -i k_0 + \o  k}=\sum_{k=h_L}^{N}
g_{\o}^{(k)}(\xx-\yy)\label{dec} \ee
and, for $h_L\le k\le N$ and for a constant $C$ (independent of $k$),
\be |g^{(k)}|_{L_1}\le C 2^{-k}, \quad\quad |g^{(k)}|_{L_\io}\le C
2^{k} \label{fon111}\ee
%
%

The decomposition of the propagator \pref{dec} allows us to make a decomposition
of the fermionic measure $P(d\psi)=\prod_{k=h_L}^N P(d\psi^{(k)})$, where $P(d\psi^{(k)})$
is the fermionic measure with propagator $g_{\o}^{(k)}(\xx)$
and
the corresponding decomposition of the field is
$\psi_{\xx,\o,j}=\sum_{k=h_L}^{N}\psi^{(k)}_{\xx,\o,j}$. Indeed
calling $\VV(\psi,J)=V(\psi)
+ \sum_{\o,j} \int \!d\xx\; J_{\xx,\o}
\psi^{+}_{\xx,\o,j} \psi^{-}_{\xx,\o,j}$ we can write the generating functional
$W_{N,L}(\h,J)$ \pref{caro} (in the $\h=0$ case for definiteness) as
\bea
&& e^{W_{N,L}(0,J)}=
\int P(d\psi)e^{\VV(\psi,J)}= \int P(d\psi^{(\le N-1)}) \int P(d\psi^{(N)}) e^{\VV(\psi^{(\le N-1)}+\psi^{(N)},J)}=\nn\\
&&e^{-L^2 E_{N-1}+S_{N-1}(J)}\int P(d\psi^{(\le N-1)}) e^{\VV^{(N-1)}(\psi^{(\le N-1)},J)}
\eea
where
\be
-L^2 E_{N-1}+S_{N-1}(J)+\VV^{(N-1)}\equiv\sum_{n=0}^\io {1\over n!}\EE^T_N(\VV;n)
\ee
and $\EE^T_N$ are the fermionic {\it truncated expectations} with propagator $g^{(N)}_\o$; the first term in the l.h.s. is a constant $J,\psi$-independent, the second depends only on the source term $J$ and the third, the {\it effective potential}
at scale $N-1$, is the sum of integrals of monomials in $\psi$ and $J$ times suitable kernels.
One could repeat the
same procedure writing $ P(d\psi^{(\le N-1)})= P(d\psi^{(\le N-2)})
 P(d\psi^{(N-1)}) $ and integrating the field $\psi^{(N-1)}$ and so on, but it turns out that this is not suitable for taking the limit $N\to\io$, the reason being that
the {\it scaling dimension} of the monomials in the effective potential with $2m$ $\psi$-fields and $n$
$J$-fields is
$D=2-n-m$, that is greater or equal to zero in the case $(2m,n)=(2,0), (4,0), (2,1)$.
These are the scaling dimensions typical of a {\it renormalizable} Quantum Field Theory and therefore one could expect that
ultraviolet divergences are present requiring that
the ultraviolet $N\to\io$ limit can be taken only with the proper choice of
the bare parameters, possibly diverging in the
$N\to\io$ limit. However, the fact that the interaction is non local but short ranged
induces an {\it improvement} in the scaling dimension, and
indeed no ultraviolet divergences are in fact present; the kernels
of the effective potential are bounded uniformly in the ultraviolet cut-off $N$.
The proof of this fact in a similar case is in \cite{M1}, and it will be recalled and extended to the present case below; this kind of analysis dates back to \cite{Le}, where it was applied to the construction of the ultraviolet limit of the Yukawa$_2$ model.

\subsection{Ultraviolet scales}

The integration procedure of the scales $N-1,...h_L$ must be different for the positive (ultraviolet) and negative (infrared) scales. We describe first the integration of the ultraviolet scales, following a procedure very
similar to the one explained in \S 2.1 and \S 2.2 of \cite{M1}.
Assume that we have integrated the Grassmann variables $\psi^{(N)},..,\psi^{(h)}$, $h\ge 1$ so obtaining
\be
e^{\WW_{N,L}(0, J)}=e^{-L^2 E_h+S_h(J)}\int P(d\psi^{(\le h)}) e^{\VV^{(h)}(\psi^{(\le h)},J)}
\label{ssssaa1}\ee
where
\bea&&
\VV^{(h)}(\psi)=\label{2.17aa}\\
&&\sum_{m=1}^\io\sum_{n=0}^\io\sum_{\underline\o,\underline
j} \int d\uz \int d\ux \int d\uy
W^{(h)}_{2m,n}(\uz;\ux,\uy) [\prod_{i=1}^n J_{\zz_i}][\prod_{i=1}^{2 m}
\psi^{\e_i}_{\xx_i, \o_i, j_i}] \nn\eea
We write the r.h.s. of \pref{ssssaa1}
 as
\be e^{-L^2 E_h+S_h(J)}\int P(d\psi^{(\le h)})e^{\LL
\VV^{(h)}(\psi^{(\le h)},J)+\RR
\VV^{(h)}(\psi^{(\le h)},J)}\label{ssss2} \ee
where $\RR=1-\LL$ and $\LL$ is a linear operation, called {\it localization operation},  acting on the
kernels so that
\be\LL W_{2m,n}^{(h)}(\uz;\ux,\uy):=
\begin{cases}
\displaystyle
W_{2m,n}^{(h)}(\uz;\ux,\uy)&\text{ if $n+m\le 2$}\\
0 &\text{ otherwise}
\end{cases}\label{loc}
\ee
Therefore we can write
\bea &&e^{-L^2 E_h+S_h(J)}\int P(d\psi^{(\le h-1)})
\int P(d\psi^{(h)})
e^{\LL
\VV^{(h)}(\psi^{(\le h)},J)+\RR
\VV^{(h)}(\psi^{(\le h)},J)}=\nn\\
&&e^{-L^2 E_{h-1}+S_{h-1}(J)}\int P(d\psi^{(\le h-1)})
e^{\VV^{(h-1)}(\psi^{(\le h-1)},J)}
\label{ssss3} \eea
and the procedure can be iterated. Note that
the localization operation simply
singles out the running coupling functions (not constants) without doing
any operation operation on them.

\subsection{Scaling dimensions}

The outcome of this integration procedure is that
the kernels of the effective potentials
are expressed in terms of an expansion in terms of running coupling functions $W^{(k)}_{2,1}, W^{(k)}_{2,0}, W^{(k)}_{4,0}$,
with $k>h$.
This expansion is described in terms of suitable {\it trees} defined
in Appendix 1  of \cite{M1} (up to trivial modifications)
and we will not repeat the details here.
We define the following norm
\be
||W^{(h)}_{2m,n}||={1\over L^2}\int d\underline \xx |W^{(h)}_{2m,n}(\xx)|\
\label{say}\ee
By Lemma 1 of \cite{M1}  (the proof is in Appendix 1 of \cite{M1}, and one can verify that the extra index $j$ and the more complex form of the interaction plays no role) it holds that, if $k>0$ and
assuming that
\be
\sup_{k>h}|| W_{2,0}^{(k)}||+||W_{4,0}^{(k)}||\le \e_0, \quad \sup_{k>h}||W_{2,1}^{(k)}||\le 2\label{xxx2}
\ee
with $\e_0$ independent from  $k,L,N$,
then for a suitable constant $C$
\be
||W^{(h)}_{2m,n}||\le C  \e_0^{\e_{2m,n}} 2^{-h(n+m-2)}\label{main}\ee
with $\e_{2m.n}=1$ for $m\not=0$ and zero otherwise.

The bound (32) establishes that the scaling dimensions of the kernels
is $2-n-m$,  which is  negative except when $(2m,n)=(2,0);(4,0);(1,2)$;
therefore the above statement
essentially says that a {\it dimensional bound} can
be proved for the kernels of the effective potential, provided that the bounds for
the kernels corresponding to
$(2m,n)=(2,0);(4,0);(1,2)$ are improved.

\subsection{Determinant bounds}

Note that the kernels $W^{(h)}_{2m,n}$ can be written as a sum over renormalized Feynman graphs; each of them is $O(2^{-h(n+m-2)})$ but their number at order $\bar n$ grows as $O(
(\bar n!)^2)$ and therefore in this way one cannot establish the convergence of the expansion.
In order to achieve convergence (and to establish \pref{main}) one needs to exploit the cancellations
due to the relative minus signs which follows from anticommutativity; technically one has to exploit the fact that the fermionic expectations can be represented in terms of Gram determinants. Therefore
the proof of \pref{main} is based on the representation of
$W^{(n;m)(k)}$ in terms of Gallavotti-Nicolo' trees, see  \cite{G} for more details, and the {\it
Brydges-Battle-Federbush} formula for the truncated expectations (see \cite{Le}, App. A), which says that
\be \EE^T_{h}(\widetilde\psi^{(h)}(P_1),\ldots,
\widetilde\psi^{(h)}(P_s))=\sum_{T}\prod_{l\in T}
\big[g^{(h)}(\xx_l-\yy_l)\big]\, \int dP_{T}({\bf t})\; {\rm
det}\, G^{h,T}({\bf t})\label{xxxxx}\ee
where $\widetilde\psi^{(h)}(P_i)=\prod_{f\in P_i}\psi^{\e(f)(h)}_{\xx(f),\o(f),j(f)}$
is the product of fields associated to the cluster $i$, $P_i^\pm$ is the set of field labels associated to $+$
or $-$ fields, that is
$P_i^\pm\=\{f\in P_i,\e(f)=\pm\}$, with $f_{ij}^\pm$
being their elements. Defining $\xx^{(i)}=\cup_{f\in P_i^-}\xx(f)$,
$\yy^{(i)}=\cup_{f\in P_i^+}\xx(f)$, $\xx_{ij}=\xx(f^-_{i,j})$,
$\yy_{ij}=\xx(f^+_{i,j})$,
$T$ is a set of lines forming an {\it anchored tree graph} between the
clusters of points $\xx^{(i)}\cup\yy^{(i)}$, that is $T$ is a set of lines,
which becomes a tree graph if one identifies all the points in the same
cluster. Moreover ${\bf t}=\{t_{ii'}\in [0,1], 1\le i,i' \le s\}$,
$dP_{T}({\bf t})$ is a probability measure with support on a set of ${\bf t}$
such that $t_{ii'}={\bf u}_i\cdot{\bf u}_{i'}$ for some family of vectors
${\bf u}_i\in \RRR^s$ of unit norm. Finally $G^{h,T}({\bf t})$ is a
$(k-s+1)\times (k-s+1)$ matrix, whose elements are given by
\be G^{h,T}_{ij,i'j'}=t_{ii'}
\,g^{(h)}(\xx_{ij}-\yy_{i'j'})\label{det} \ee
with $(f^-_{ij}, f^+_{i'j'})$ not belonging to $T$.
Finally a couple
$l\=(f^-_{ij},f^+_{i'j'})\=(f^-_l,f^+_l)$ will be called a line joining the
fields with labels $f^-_{ij},f^+_{i'j'}$
and connecting the points
$\xx_l\=\xx_{i,j}$ and $\yy_l\=\yy_{i'j'}$, the {\it endpoints} of $l$.

Let $\HH=\RRR^s\otimes \HH_0$, where $\HH_0$ is the Hilbert space of complex
vectors $F(\kk)$ with scalar product
\be
<F,G>={1\over L^2}\sum_{\kk} F^*(\kk) G(\kk)\;.
\ee
Note that
\pref{det} can be written in the form, see \cite{Le} App. A
\be
G^{h,T}_{ij,i'j'}=t_{ii'}
\,g^{(h)}(\xx_{ij}-\yy_{i'j'})=<\uu_i\otimes A^{(h)}_{\xx(f^-_{ij})},
\uu_{i'}\otimes B^{(h)}_{\xx(f^+_{i'j'})}>
\ee
where $\uu_i\in \RRR^s$, $i=1,\ldots,s$, are the vectors such that
$t_{i,i'}=\uu_i\cdot\uu_{i'}$ and (the cut-off function $f_h(\kk)$
should not be confused the index $f$ defined after (33))
\bea A_h(\xx)&=& \sqrt{f_{h}(\kk)}
\frac{e^{i\kk\xx}}{ k_0^2+ k^2}\;,
\nn\\
B_h(\xx)&=& e^{i\kk\xx}\sqrt{f_h(\kk)} (ik_0+\o  k)\label{g1}\eea
One can use the well known {\it Gram-Hadamard inequality}, which states
that, if $M$ is a square matrix with elements $M_{ij}$ of the form
$M_{ij}=<A_i,B_j>$, where $A_i$, $B_j$ are vectors in a Hilbert space with
scalar product $<\cdot,\cdot>$, then
\be|\det M|\le \prod_i ||A_i||\cdot ||B_i||\;.\ee
where $||\cdot||$ is the norm induced by the scalar product. Therefore one can use
the above inequality, and the fact that
\be ||A^{(h)}(\xx)||^2\le C 2^{-2h}\;,\quad\quad
||B^{(h)}(\xx)||^2\le C 2^{4h}\label{g2}\ee
for a suitable constant $C$, to bound the determinants in the truncated expectations without dangerous factorials. One has also to take into account the tree structure induced by the multiscale analysis and the
effect of the $\RR$ operation; we refer to App. 1 of \cite{M1} for the complete proof, up to simple modifications.

\subsection{Improvement of scaling dimension}

It remains to prove \pref{xxx2}. That is, we must show that there is an improvement in the scaling dimension
of the kernels with non negative dimension; for the single chain model this
was the content of Lemma 2 of \cite{M1} whose
proof is in \S 2.4-2.6 of \cite{M1} (see also \S IV of \cite{Le}) and we have to check that the analysis can be extended also to the present more complicated form of the interaction.

One assumes that
\pref{xxx2} is true for $h+1$ (so that the bound \pref{main} holds) and then one proves it for the scale $h$. Denoting by $\EE^T_{h,N}$ the fermionic truncated expectation
with respect to $g^{(h,N)}$, we recall the following property, (see {\it e.g.} eq.(15) in \cite{Le})
\bea
&&\EE^T_{h,N}(\tilde\psi(P_1\cup P_2)\tilde\psi(P_3)...\tilde\psi(P_n))=\EE^T_{h,N}(\tilde\psi(P_1)\tilde\psi(P_2)...\tilde\psi(P_n))\label{sasa1}\\
&&
+\sum_{K_1,K_2\atop K_1\cup K_2=\{3,..,n\}, K_1\cap K_2=\{0\}}(-1)^\pi\EE^T_{h,N}(\tilde\psi(P_1)\prod_{j\in K_1}\tilde\psi(P_{j}))
\EE^T_{h,N}
(\tilde\psi(P_2)\prod_{j\in K_2}\tilde\psi(P_{j}))\nn
\eea
where $(-1)^\pi$ is the parity permutation necessary to bring the Grassmann variables in the r.h.s. in the original order.
By using \pref{sasa1} we can decompose the kernel $W_{2,0}^{(h)}$
as in Fig. 1 (more details can be found in \S 2.3 of \cite{M1}) and we get the bound
\be
|W_{2,0}^{(h)}(\xx_1,0)|\le
C\max |\vec g|] \int d\xx_2 d\xx_3 |v(\xx_1-\xx_2)
g^{(h,N)}(\xx_1-\xx_3) W_{2,1}^{(h)}(\xx_2;\xx_3,0)|\label{sasa4}
\ee
where we have taken into account that
the first and last term in Fig. 1 are vanishing as $g^{(h,N)}(\kk)=-g^{(h,N)}(-\kk)$.

\insertplot{100}{85}
{\ins{130pt}{60pt}{$+$}
\ins{50pt}{60pt}{$=$}
\ins{280pt}{60pt}{$+$}}
{verticiT11}
{\label{h2} Graphical representation of the decomposition of the
kernel $W_{2,0}^{(h)}$; ; the blobs represent $W^{(h)}_{2m,n}$, the
paired wiggly lines represent $v$, the full lines $g^{(h,N)}$ and
the dotted lines are the external fields} {0}

%

A naive bound for $||W_{2,0}^{(h)}||$ is
\bea &&[\max |\vec g|]||\int d\xx_1 d\xx_2 d\xx_3 v(\xx_1-\xx_2)
g^{(h,N)}(\xx_1-\xx_3) W_{2,1}^{(h)}(\xx_2;\xx_3,0)|\le\nn\\
&&[\max |\vec g|]||g^{(h,N)}|_{L_\io} |v|_{L_1} \int d\xx_2 d\xx_3
|W_{2,1}^{(h)}(\xx_2;\xx_3,0)|\le C 2^{N}, \eea
which is diverging as $N\to\io$; one can read in the exponent of
the r.h.s. that the scaling dimension is $+1$.
However one can
improve such estimate noting that the wiggly line is not necessary
to ensure the connectivity of the diagram, and one can instead integrate
over the fermionic propagator getting the bound
\bea &&[\max |\vec g|]|\int d\xx_1 d\xx_2 d\xx_3 |v(\xx_1-\xx_2)|
g^{(h,N)}(\xx_1-\xx_3) W_{2,1}^{(h)}(\xx_2;\xx_3,0)|\le\nn\\
&&[\max |\vec g|]|g^{(h,N)}|_{L_1} |v|_{L_\io} \int d\xx_2 d\xx_3
|W_{2,1}^{(h)}(\xx_2;\xx_3,0)|\le C [\max |\vec g|] 2^{-k}\label{say2}\eea
which verifies, assuming $\max |\vec g|$ small enough, \pref{xxx2}. One can read from \pref{say2}
that the effective dimension is $-1$ instead of the previous $+1$, so that
such a contribution now behaves as an irrelevant term. In the above
bound a crucial role is played by the fact that the potential is short ranged and non local; with a local delta-like interaction $|v|_{L_\io}$ is unbounded. Note also that no dangerous factorials are generated in these bounds,
since the determinants in \pref{xxxxx} are preserved.

Similar considerations can be done for the other terms,  like
$W_{2, 1}^{(h)}$ (for more details in a similar case, see \S 2.4 of \cite{M1}). It is particularly interesting here to focus on the subset of the contribution which is identical to
$W_{0,2}^{(h)}$, which can be decomposed as shown in Fig. 2, up to other vanishing terms.
The second term in Fig. 2 can
again be bounded by
\bea && C [\max |\vec g|]  |v|_{L^\io} |W^{(h)}_{2,2}|_{L^1}\sum_{h\le i'\le
j\le i\le N}|g^{(j)}|_{L^1} |g^{(i)}|_{L^1}|g^{(i')}|_{L^\io}\le\nn\\
&& C_1 [\max |\vec g|]|^2 2^{-2h}\sum_{h\le i\le N} (i-h)2^{-i+h}\le C_2
 [\max |\vec g|]^2 2^{-2h}. \eea

\insertplot{700}{100} {
\ins{150pt}{60pt}{$+$}}
{verticiT15x}
{\label{h4} Decomposition of $W^{(h)}_{0,2}$: the blobs represent
$W^{(k)}_{2m,n}$, the paired wiggly lines represent $v$, the paired
line $g^{(h,N)}$ }{0}

Regarding the first term in Fig. 2, the possible contributions
to the bubble term
are vanishing at zero external momentum; indeed if
$\chi_{h,N}(\kk)=\sum_{i=h}^N f_i(\kk)$ then
\be {1\over L^2}\sum_\kk {\chi_{h,N}(\kk)\over (-ik_0+ k)^2}=
{1\over L^2}\sum_\kk \chi_{h,N}(\kk){k_0^2- k^2+2 i k_0
k\over (k_0^2+ k^2)^2}= 0. \label{xxx} \ee
In this way the bound in the first term has an extra $2^{-h}$ factor.
Note the crucial role played by the fact that the interaction is invariant under the transformation
\pref{zzz}; if such symmetry is broken, for instance by the
present of terms of the form $\int d\xx d\yy v(\xx-\yy)\psi^+_{\xx,\o,j}
\psi^-_{\xx,-\o,j})\psi^+_{\yy,-\o,j}
\psi^-_{\xx,\o,j}$, then the bubble graph gives a contribution $O(\log N)$.
This concludes the sketch of the proof of the bound \pref{say} (for the complete proof in a similar case, see \S 2 of \cite{M1}).

\subsection{Infrared scales}

It remains to discuss the
integration of the infrared negative scales. Now there is a crucial difference with respect to the analysis
in the effective single chain model considered in \cite{M1}; in that case, see \S 2.7 of \cite{M1},  one can perform a Renormalization Group analysis of the infrared problem and, thanks to suitable cancellations in the flow equations
 \cite{BM0}, \cite{BM1},\cite{BM2} implying the boundedness of the running coupling constants, one gets results {\it uniform} in $L$, and the thermodynamic limit can be taken. A similar analysis for the effective two chains model apparently gives instead an {\it unbounded} flow for the running coupling constant in the infrared.
However, since we are assuming $L$ finite, we
can integrate
directly over the variable $\psi^{(h_L,0)}$. Note that
\be |g^{(h_L,0)}|_{L_1}\le C L, \quad\quad |g^{(h_L,0)}|_{L_\io}\le C
\label{fon1111}\ee
Moreover the propagator has still
a Gram representation
\be
g^{(h_L,0)}(\xx-\yy)=< A(\xx), B(\yy)>\label{xxx3}
\ee
with
\bea A(\xx)&=&e^{i\kk\xx}\sqrt{\chi_{h_L,0}(\kk)}
\frac{1}{ k_0^2+ k^2}\;,
\nn\\
B(\xx)&=& e^{i\kk\xx}\sqrt{\chi_{h_L,0}(\kk)} (ik_0+\o  k)\eea
and
\be
||A||^2\le C L^2\quad ||B||^2\le C
\ee
It is an immediate consequence of \pref{xxxxx} and \pref{xxx3} and of
the Gram inequality that the contribution of order $\bar n$ to
$W^{(h_L-1)}_{2m,n}$
is bounded by $(C  [\max |\vec g|] L)^{\bar n}$,
%
%
so that we have convergence for $|\vec g|\le C L^{-1}$. With some more effort
and using a multiscale integration also for the infrared region one can arrive at
$|\vec g|$ of order
$O((\log L)^{-\k})$
for some positive $\k$.  For getting results uniform in $L$ new ideas seem necessary.

\subsection{Proof of Ward Identities}

Let us finally consider now the correction term $\hat\D_{L,N}(\kk,\kk+\qq)$
appearing in the WI \pref{h1a} in order to prove \pref{ll22}: again the proof is similar
to the one explained
in \S 3 of \cite{M1} but with some important differences.
As a first step we introduce the generating functional
\be e^{W^R_{N,L}(\h,J)} = \int\!
P(d\psi)\; e^{ -V(\psi)
+ A_0(\chi,\psi)-A_1(\chi,\psi)+ \sum_{\o,j} \int\! d\xx\; \lft[\psi^{+}_{\xx,\o,j}
\h^-_{\xx,\o,j} + \h^+_{\xx,\o,j}
\psi^{-}_{\xx,\o,j}\rgt] }\;\label{caro1}\ee
with
\bea &&A_0(\chi,\psi)=\sum_{j,\o} \int d\kk d\qq \chi_{\qq,\o,j}C_{\o}^N(\kk,\qq)
\hat\psi^+_{\kk+\qq,\o,j}\hat\psi^-_{\kk,\o,j}\nn\\
&&A_1(\chi,\psi)=\sum_{j,\o} \int d\kk d\qq \chi_{\qq,\o,j}[ {g_0\over 4\pi } D_{-\o}(\qq)\hat v(\qq)
\hat\psi^+_{\kk+\qq,-\o,j}\hat\psi^-_{\kk,-\o,j}
+{g_f\over 4\pi} D_{-\o}(\qq)\hat v(\qq)
\hat\psi^+_{\kk+\qq,-\o,-j}\hat\psi^-_{\kk,-\o,-j}\nn\\
&&+{g_4\over 4\pi} D_{-\o}(\qq)\hat
v(\qq)
\hat\psi^+_{\kk+\qq,\o,-j}\hat\psi^-_{\kk,\o,-j}+{\tilde g_4\over 4\pi } D_{-\o}(\qq)\hat
 v(\qq) \hat\psi^+_{\kk+\qq,\o,j}\hat\psi^-_{\kk,\o,j}]
\label{fondddw} \eea
where $C_{\o}^N(\kk,\qq)$ was defined in \pref{fonddd1}
and again we have used the convention $-a=b, -b=a$. This is the generating
functional for $\D_{L,N}$ minus the expected limit.
We define
\be
\hat R_{N,L}(\kk,\kk+\qq)={\partial^{3}W^R_{N,L}(\h,\chi)
\over\partial \hat\chi_{\qq,j',\o'}\partial\hat\h^{+}_{\kk,j,\o}
\partial\hat\h^{-}_{\kk+\qq,j,\o}}\Big |_{0,0}
\ee
and the proof of (19) consists in showing that, for $\kk,\kk+\qq$
fixed
\be
\lim_{N\to\io} \hat R_{L,N}(\kk,\kk+\qq)
=0\label{fro}
\ee
One can analyze $W^R_{N,L}(\h,\chi)$ with a multiscale integration procedure similar to the one described above for the generating functional (see \S 3.1 and \S 3.2 of \cite{M1} for more details in a similar case). We can integrate $\psi^{(N)},..,\psi^{(h)}$, $h\ge 1$ obtaining
\be
e^{\WW^R_{N,L}(0, \chi)}=e^{-L^2 \tilde E_h+S_h(\chi)}\int P(d\psi^{(\le h)}) e^{\tilde\VV^{(h)}(\psi^{(\le h)},\chi)}
\label{ssssaa}\ee
where $\tilde\VV^{(h)}(\psi^{(\le h)},\chi)$
has an expression similar to \pref{2.17aa}
with $\chi$ replacing $J$ and
kernels $\tilde W^{(k)}_{n,2m}$ (of course
$\tilde W^{(k)}_{2m,0}=W^{(k)}_{2m,0}$).
In the multiscale integration one
defines a localization operation extracting the
terms with non negative
dimension, which are $\tilde W^{(k)}_{2,0},\tilde  W^{(k)}_{4,0}$ and $\tilde W^{(k)}_{2,1}$;
the first two obey the bound \pref{xxx2}; the last one obeys
\be
\sup_{N>k>0}||\tilde W_{2,1}^{(k)}||\le C  [\max |\vec g||]2^{-N/2}\label{xxx22}
\ee
that it is vanishing as $N\to\io$. The proof is similar to the one in \S 3.3 of \cite{M1}.
One writes
$\tilde W_{2;1}^{(k)}=\tilde W_{2,1;a}^{(k)}+\tilde W_{2;1;b}^{(k)}$, where
$\tilde W_{2,1;a}^{(k)}$ is obtained
contracting  $A_0$ and $\tilde W_{2;1;b}^{(k)}$ contracting $A_1$. In order
to study the properties of $\tilde W_{2,1;a}^{(k)}$ we further distinguish
the case in which one of the external lines belongs to $A_0$ or not;
in the first case the bound \pref{xxx2} follows easily by a parity cancellation
and the fact that the contracted line is necessarily at scale $N$.
It remains then to consider the case in which both the fields in $\AA_0$ are contracted. An important role is played by the function
\be
C^N_\o(\kk,\qq)\hat g^{(h)}_{\o}(\kk) \hat g^{(i)}_{o}(\kk+\qq)=0\quad\quad
h,i<N\label{man}
\ee
We can decompose such class of terms as we did for
$W_{2,1;a}^{(k)}$, see fig. 3.
\insertplot{700}{90}
{\ins{150pt}{60pt}{$+$}}
{verticiT15xx}
{\label{m9} Contributions to $\tilde W^{(k)}_{2,1}$; the black dot
represents $A_0(\chi,\psi)$} {0}
The second term in
Fig. 3 can be bounded by
\bea &&[\max |\vec g|]  |v|_{L^\io} |\tilde W^{(k)}_{2,2}|_{L^1}\sum_{k\le
i'\le i\le N}|g^{(N)}|_{L^1}
|g^{(i)}|_{L^1}|g^{(i')}|_{L^\io}\le\nn\\
&& C [\max |\vec g|] ^2 2^{-2k}(N-k)2^{-N+k}\le {\rm Const.}\l^2
2^{-2k}2^{-(N-k)/2}, \eea
leading to the vanishing of this contribution for $N\to\io$
together with a negative dimension; with respect to (44), the scale $(j)$ is not summed but blocked at scale $N$ by \pref{man}.

It now remains to consider the contribution from the first term in Fig. 3, which is the only possible non vanishing one.
Here it comes the main differences with respect to the case discussed in \cite{M1}.
First of all, the possibility of this kind of term to contribute depends crucially on the form of the interaction; in particular,
there is {\it no contribution} of this kind involving $g_u$ and $g_{bs}$ (this explains why the anomaly does not depend from such couplings). Moreover, the contributions to the bubble coming from the other couplings are {\it non} vanishing; the value of the bubble in Fig. 3 is given by
\bea &&{1\over L^2}\sum_{\kk}{C^N_{\o}(\kk,\qq)\over
D_{-\o}(\qq)} g_{\o}(\kk)g_{\o}(\kk+\qq)=
\label{bub}\\
&&={1\over L^2} \sum_{\kk} { \chi_N(\kk + \qq)-\chi_N(\kk
)\over D_{-\o}(\qq) D_\o(\kk+\qq)}
 + {1\over
L^2}\sum_{\kk}{(\chi_N(\kk + \qq)- 1)\chi_N(\kk)\over
D_{\o}(\kk) D_{\o}(\kk+\qq)}.\nn \eea
The second line can be rewritten, changing the variables
\be{1\over L^2 2^{2N}} \sum_{\kk'} {  \chi_0(\kk' + \qq 2^{-N})-\chi_0(\kk')\over D_{-\o}(\qq 2^{-N}) D_{\o}(\kk+\qq 2^{-N})}
 + {1\over
L^2 2^{2N}}\sum_{\kk'}{(\chi_0(\kk' + \qq 2^{-N})- 1)\chi_0(\kk')\over
D_{\o}(\kk') D_{\o}(\kk'+\qq 2^{-N})}, \label{36}\ee
where $\kk'=2^{-N}\kk$. As a result of that the second term can be written as
\be {1\over L^2 2^{2N}}\sum_{\kk'}{(\chi_0(\kk')- 1)\chi_0(\kk')\over
D_{\o}(\kk') D_{\o}(\kk')}+O(\qq 2^{-N})\label{37} \ee
and the first term in \pref{37} vanishes by symmetry. Moreover,
the first term in \pref{36} can be rewritten in the limit
$N\to\io$ as
\be -\int_0^\io d\r \int_0^{2\pi} d\th\partial_\r
\bar\chi(\r){\sin^2\th\over 4\pi^2} =-{1\over 4\pi}\int_0^\io d\r
\partial_\r \bar\chi(\r) ={1\over 4 \pi }. \ee
and is {\it independent} from $L$.
Such terms are therefore canceled by
the corresponding contributions coming from $\tilde W_{2,1;b}^{(k)}$.
Note also the extra factor
$[\max |\vec g||]$ in \pref{xxx22} as from \pref{man} there is no contribution
of order $0$ at scales $<N$. Therefore from \pref{xxx22}
we get, for $n\not=0$  and $h>0$
\be
||\tilde W^{(h)}_{m,2n}||\le C  2^{-h(n+m-2)}2^{-N/2}\label{xx}\ee
%
%
%
from which \pref{fro} follows.
\vskip3cm

\section{Conclusions}

We have considered the effective two chain model \pref{caro}
described in terms of bonding and antibonding fields with linear
dispersion relations. Contrary to the effective single chain problem, the
model is not exactly solvable and the associated WI have never been derived
by bosonization methods. By analyzing the functional
integrals using Constructive Quantum Field Theory methods we have
derived for the first time exact WI in the limit of removed ultraviolet cut-off (and at finite volume) which
displays the presence of chiral anomalies. Such
anomalies do not depend directly on the umklapp and backscattering
interactions and verify the Adler-Bardeen non renormalization property.
Such a WI provide non trivial relations between correlation
functions and are likely to play an important role
for the understanding of the
thermodynamic limit of the model \pref{caro} which are still largely unknown.

\section{Appendix A}

The WI \pref{h1a},(19) has been derived assuming an ultraviolet cut-off
function $\chi_N(\kk)$ in the propagator
\pref{pro}, cutting the momenta (that is, spatial momenta and energies)
greater than $|\kk|\ge 2^{N+1}$.
While the existence of the chiral anomaly does not depend
on the choice of the cut-off, some of its properties may depend on it.
It is therefore interesting to consider a (somewhat more realistic) cut-off only
on the spatial component of the momentum.
We consider \pref{caro} with an integration
 $P(d\psi)$ whose propagator
is not \pref{pro} but
\be g_{\o}(\xx-\yy)={1\over
L^2}\sum_{\kk\in \DD}e^{i\kk(\xx-\yy)}{\chi_{N}(k)\over -i k_0+\o v_F k}\label{pro1} \ee
with $\chi_N(k)=\bar\chi(2^{-N} v_F |k|)$, that is the cut-off depends only from the spatial part
of the momentum; for later convenience we introduce also the Fermi velocity $v_F$ which was set equal to $1$ before. Note that this cut-off breaks the Lorentz invariance of the theory.

An analysis similar to the previous one leads to
a WI similar to \pref{h1a} with $D_\o(\kk)=-i k_0+\o v_F k$; moreover the correction
$\hat \D_N$ verifies
\pref{h3} with
$D_{\o}(\qq)$ replaced by
$2 \o q$. Note indeed that in the computation
of the bubble \pref{bub} the first term in Fig. 3 is now given by
\be \lim_{N\to\io}{1\over L^2 2^{2N}} \sum_{\kk'} { \chi_0(k' +
q 2^{-N} )-\chi_0(k' )\over (-\o  q v_F 2^{-N})
D_{\o}(\kk'+
q 2^{-N})}=\int {d\kk'\over (2\pi)^2} {\partial \chi_0(k')
\over -\o D_{\o}(\kk')} ={1\over 2\pi v_F }\ee
while the second term is still vanishing. Choosing the cut-off function $\chi_N(\kk)$ the anomalous term
(19) is Lorentz invariant while with  a cut-off  $\chi_N(k)$
Lorentz invariance is never restored in the WI even in the limit $N\to\io$.

It is also interesting to write the WI for the total densities for
cut-off $\chi_N(k)$; one gets in the limit $N\to\io$
\bea &&\sum_j (-i q_0+\o (v_F-{g_4+\tilde g_4\over 2\pi}) q
)<\hat\r_{\qq,\o,j}\hat\r_{-\qq,\o',j'}>+\label{26}\\
&&\sum_j{1\over 2\pi}(g_0+g_f)\o q
<\hat\r_{\qq,-\o,j}\hat\r_{-\qq,\o',j'}>=\d_{\o,\o'}\o q. \nn \eea
If $\hat \r^\r_{\qq}=\sum_{j,\o} \hat \r_{\qq,\o,j}$ is the total density,
we can derive from \pref{26} the
explicit form of the total density correlation:
%
%
%
\be <\hat\r^\r_{\qq}\hat\r^\r_{-\qq}>=K {q^2\over q_0^2+v_\r^2
q^2}+O(\qq)\label{ss}\ee
where
\be v_\r=\sqrt{ \left( v_F-{g_4+\tilde g_4\over 2\pi} \right)^2 - \left( {g_0+g_f\over
2\pi} \right)^2}, \quad\quad  K=1-{g_0+g_f\over 2\pi}. \ee
%
%
%

Finally we note that
the choice of the cut-off $\chi_N(k)$ only on the spatial momenta and the assumption
that $g_{u}=g_{bs}=0$ and of a potential
$\hat v(p)$ depending only on the
space momenta makes the model \pref{caro}
very closely related to the (spinning) Luttinger model.
It was shown in \cite{ML} that the Hamiltonian of such a model
can be expressed as a quadratic Hamiltonian in terms of the fermionic densities
$\r_{\o,j}(p)$ verifying the following commutations rules
\be [\r_{\o,j}(p),\r_{\o',j'}(-p')]=\d_{\o,\o'}\d_{j,j'} {p L\over
2\pi}. \ee
and as the Hamiltonian is bilinear in the
boson fields one gets
\bea &&{\partial \r_{\o,j}(p)\over \partial t}=[H,\r_{\o,j}(p)]=\o v_F
p \r_{\o,j}(p)+\o {p L\over 2\pi} g_0 v(p) \r_{-\o,j}(p)+\nn\\
&& \o {p L\over 2\pi} g_f v(p) \r_{-\o,-j}(p)+\o {p L\over 2\pi}
g_4 \hat v(p) \r_{\o,-j}(p)+\o {p L\over 2\pi} g_4 \hat v(p) \r_{\o,j}(p)\label{xxx}
\eea
From the above relations one gets, in the Luttinger model,
a WI identical to \pref{h1a},\pref{h3}
with
$D_{\o}(\qq)$ replaced by
$2 \o  q$, in agreement with was is found in \pref{caro}
with a cut-off $\chi_N(k)$.

A crucial point to be stressed is that, although the bosonization method was used
only in the $g_{u}=g_{bs}=0$ case, the functional integral derivation explained in this paper successfully deals
with this more general case.
There were indeed attempts to apply bosonization also
in presence of $g_u$ and $g_{bs}$, but the bosonic expression for such terms
is known to be highly formal and the conclusions drawn from it are
not really safe and unambiguous, as was stressed long ago in \cite{SO79}.

\section{Appendix B}

It is useful, for readers not familiar with Constructive Renormalization Group methods,
to verify the WI \pref{h1a},(19)
by a perturbative computation up to two loops, which is the minimal order
at which the anomaly non renormalization can be tested.
We assume as in App. A
the cut-off only for spatial momenta and the propagator \pref{pro1};
the WI are therefore given, as explained in App. A, by (in terms of amputated correlations)
\bea
\sum_j (i q_0 &-& \o'  v_F q) \Lambda_{\o, {\o}'}^{j,
j'} ({\bf k+q}, {\bf k}) -
\sum_j \o' v_F q [ ( {\bar g}_0 + {\bar g}_f ) \Gamma_{\o, {- \o}' }^{j, j'} ( {\bf k+q}, {\bf k} )  + \nonumber \\
&+&  ( {\bar g}_4 + {\bar {\tilde g}}_4 ) \Gamma_{{\alpha},
{\o}' }^{j, j'} ( {\bf k+q}, {\bf k} ) ] = \delta_{ {\o}',
\o} [ \Sigma_{{\o}}^{j} ( {\bf k} ) - \Sigma_{
{\o}}^{j} ( {\bf k+q} )  ] \label{ss}\eea
where as usual $\G$ and $\Sigma$ are, respectively,  the vertex part and the self-energy,
${\bar g}_i = g_i / 2 \pi v_F$ and
\be \Gamma_{\o, {\o}'}^{j, j'} ({\bf p+q}, {\bf p}) = \delta_{\o, {\o}'} + \Lambda_{\o,
{\o}' }^{j, j'} ({\bf p+q}, {\bf p}). \ee

Let us choose ${\o}= + $ and $ j= b $. In this way, in
1-loop order equation becomes
\bea (i q_0 &-& v_F q) [ \Lambda_{+,+}^{b,b \; (1)} ({\bf k+q},
{\bf k}) + \Lambda_{+,+}^{b,a \; (1)} ({\bf k+q}, {\bf k}) ]
- v_F q  ( {\bar g}_4 + {\bar {\tilde g}}_4 ) = \nonumber \\
&=& \Sigma_{+}^{b \; (1)}({\bf k}) - \Sigma_{+}^{b \; (1)}({\bf
k+q}),\label{an}\eea
for $ \o' = + $, and
\be (i q_0 + v_F q) [
\Lambda_{+,-}^{b,b \; (1)} ({\bf k+q}, {\bf k}) +
\Lambda_{+,-}^{b,a \; (1)} ({\bf k+q}, {\bf k}) ] + v_F q  ( {\bar
g}_0 + {\bar g}_f ) = 0\label{an1} \ee
for $ \o' = - $. In our notation, $ \Lambda_{\o {\o}'}^{j, j' \; (1)} $
represents the corresponding 1-loop vertex function.
The Feynman graphs contributing to \pref{an} are represented in Fig. 4; the Feynman
graph (c) contributing to the vertex part is matched exactly by the corresponding self-energy
difference in 1-loop order. In contrast the diagrams (a) and (b) contributing to the vertex
part, computed with a cut-off on spatial momenta to be removed at the end, are canceled
exactly by the anomalies (first term in the second line of Fig. 4). Indeed the contribution
to $\hat\D_N$ at this order is given by a graph equal to (a) and (b) with the wiggly line connected
to a black dot as in Fig 3, and one can check that such graphs are equal to $v_F(\bar g_4 + ¯
\bar{\tilde g_4})$ when
the momentum cut-off is removed. Similarly the Feynman graphs contributing to \pref{an1} are
represented diagrammatically in Fig. 5. Again the vertex diagrams are canceled entirely by
the corresponding anomalous terms produced by $g_0$ and ¯
$g_F$.

\begin{figure}[h]
\includegraphics[height=1.7in, width=4.7in]{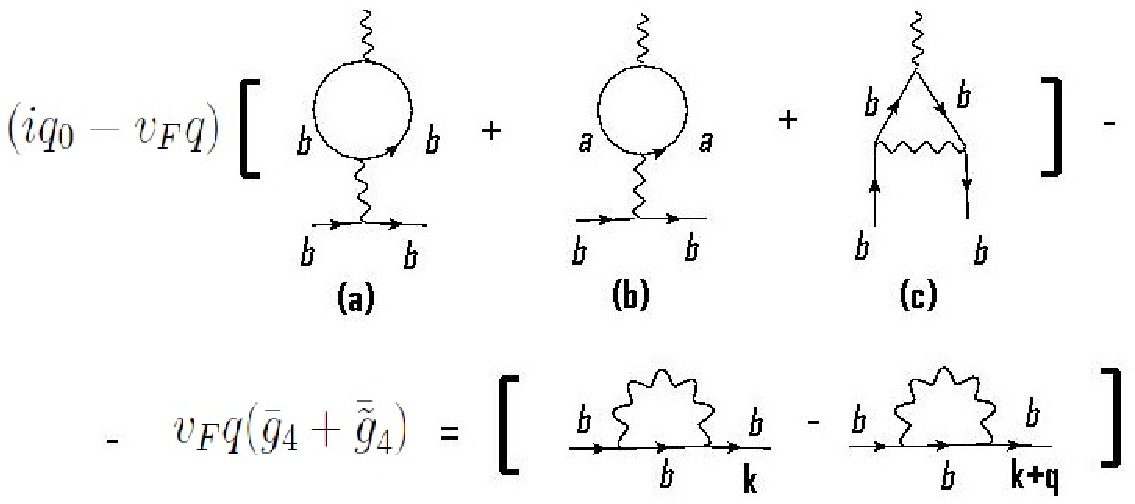}
\caption{Feynman graphs contributing to \pref{an}.}
\end{figure}

\begin{figure}
\includegraphics[height=0.9in, width=4.3in]{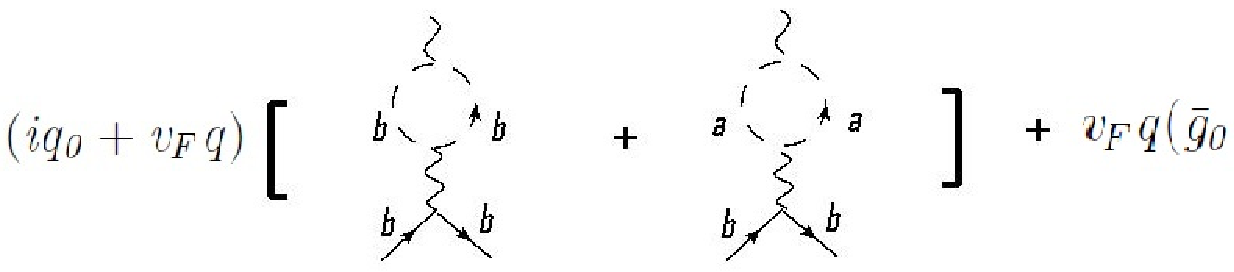}
\caption{Feynman graphs contributing to \pref{an1}}
\end{figure}

At second order the WI \pref{h1a},\pref{h3} is given by
\bea
(i q_0 &-& v_F q) \sum_{j= b, a} \Lambda_{+,+}^{b,j \; (2)} ({\bf
k+q}, {\bf k}) -
v_F q  \sum_{j=b,a} [ ( {\bar g}_0 + {\bar g}_f ) \Lambda_{+,-}^{b, j \; (1)} ( {\bf k+q}, {\bf k} )  + \label{an3} \\
&+&  ( {\bar g}_4 + {\bar {\tilde g}}_4 ) \Lambda_{+,+}^{b,j \;
(1)} ( {\bf k+q}, {\bf k} ) ] = \Sigma_{+}^{b \; (2)} ( {\bf k} )
- \Sigma_{+}^{b \; (2)} ( {\bf k+q} )  ] \nn \eea and \bea (i q_0
&+& v_F q) \sum_{j= b, a} \Lambda_{+,-}^{b,j \; (2)} ({\bf k+q},
{\bf k}) +
v_F q  \sum_{j=b,a} [ ( {\bar g}_0 + {\bar g}_f ) \Lambda_{+,+}^{b, j \; (1)} ( {\bf k+q}, {\bf k} )  + \label{an4} \\
&+& ( {\bar g}_4 + {\bar {\tilde g}}_4 ) \Lambda_{+,-}^{b,j \;
(1)} ( {\bf k+q}, {\bf k} ) ] = 0\nn \eea



We display in Fig 6. the graphs contributing to (63). The diagrams (2), (3), (4), and
diagrams (14) and (15) cancel each other out exacly. The diagrams (7),(8),(9) are canceled by the
second contribution to the self-energy difference. Similarly  (10), (11), (12)
are canceled by the first contribution,
(13) by the third and so on. The Feynman diagrams (1), (5), (6) are canceled by the
anomaly contributions given by the fourth line in Fig 6 (coming from diagrams of the form
(1),(5),(6) in which the external wiggly line is connected to a black dot as in Fig. 3).

\newpage

\begin{figure}[h]
\includegraphics[height=3.8in, width=4.9in]{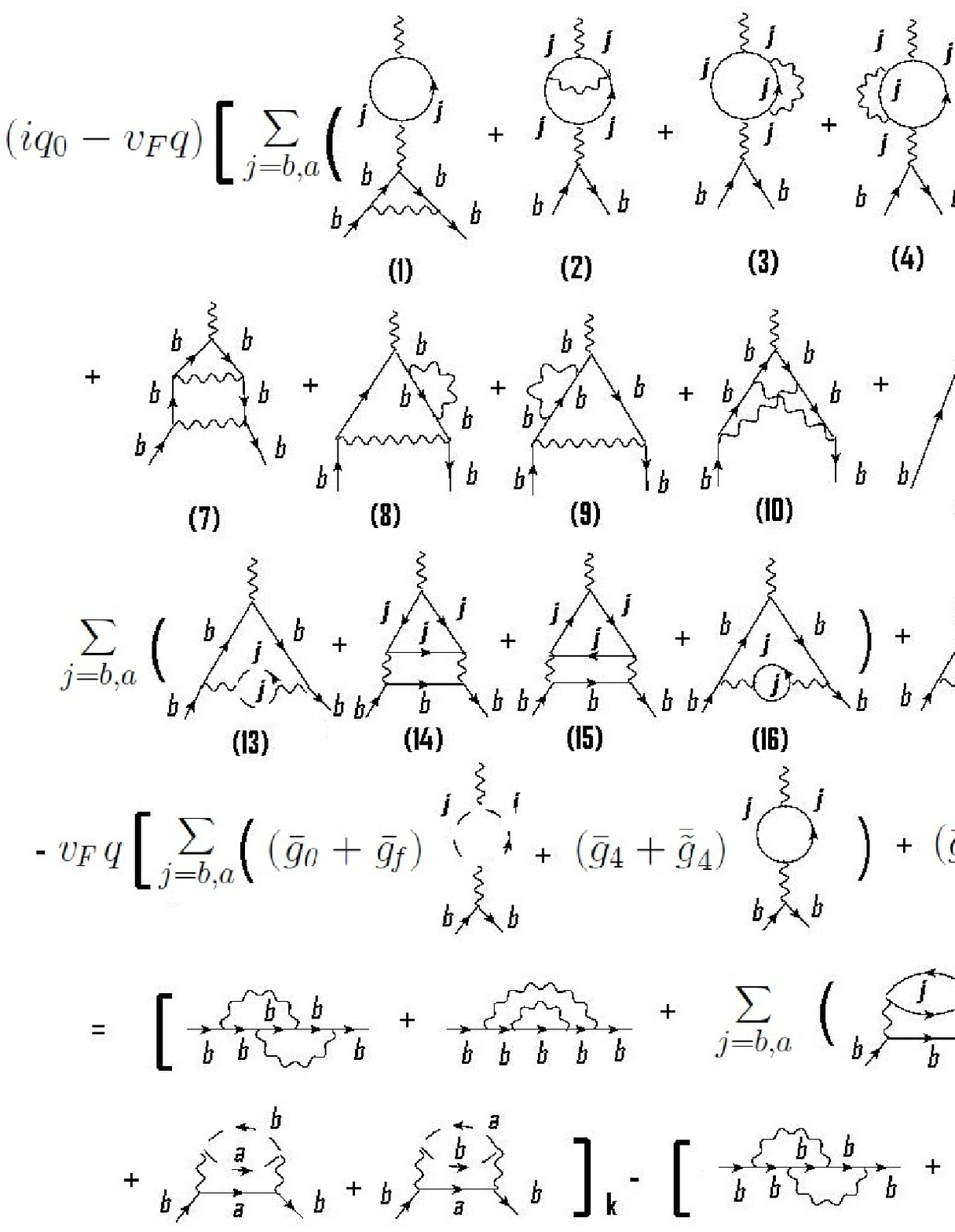}
\caption{Feynman graphs contributing to \pref{an3}}
\end{figure}

\begin{figure}[h]
\includegraphics[height=2.8in, width=5.3in]{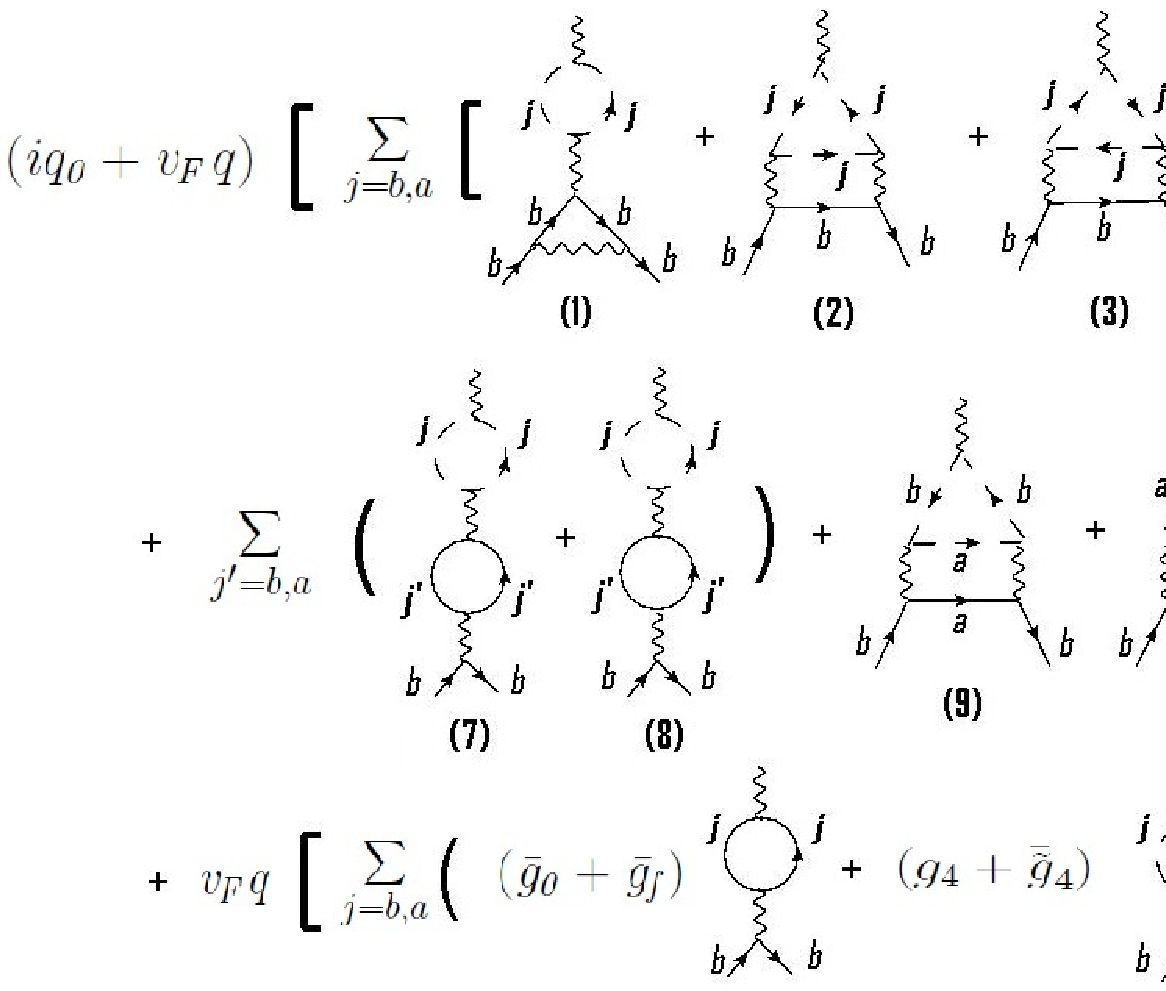}
\caption{Feynman graphs contributing to \pref{an4}}
\end{figure}

Similarly we display (64) in Fig 7; the diagrams (1), (7),(8) cancel with the anomalous
terms, and the vertex diagrams (2) and (3), (9) and (10), (11) and (12) as well as diagrams
(4), (5) and (6) cancel each other out exactly. One clearly sees that all anomalous diagrams
are associated with the $g_0$, $g_f$ and $g_4$,
couplings which are originated in 1-loop order. In
other words there are no new classes of anomalous vertex diagrams in higher loops so that
the anomaly is linear in the coupling and the $g_u$, $g_{bs}$ couplings contribute in many ways to both the vertex
and the self-energy but not to the anomaly, in agreement with the Theorem proved in this paper.


\vskip.8cm
{\rm Acknowledgments} {\it
L.C. and A.F. thanks
the CNPq, CAPES (Probral) and the Brazilian Min. of Science and Technology; V.M.
acknowledge financial support from the
ERC Starting Grant CoMBoS-239694 and from MIUR.}

%
%


\end{document}